# Security Mental Model: Cognitive map approach

Tahani Albalawi, Kambiz Ghazinour, and Austin Melton
Department of Computer Science
Kent State University
Kent, USA
{talbala1, kghazino, amelton}@kent.edu

*Abstract*— Security models have been designed to ensure data is accessed and used in proper manner according to the security policies. Unfortunately, human role in designing security models has been ignored. Human behavior relates to many security breaches and plays a significant part in many security situations. In this paper, we study users' decision-making toward security and usability through the mental model approach. To elicit and depict users' security and usability mental models, crowd sourcing techniques and a cognitive map method are applied and we have performed an experiment to evaluate our findings using Amazon MTurk.

*Keywords—mental model; cognative map; security; usability; human factor; decision; MTurk*

## I. INTRODUCTION

The computer security community has developed formal methods for providing security properties to systems and organizations. However, the human role has often been overlooked in security. How human behavior relates to many security breaches and incidents has only recently been recognized [20]. Expanding on this concept, [19] shows that security is not only a mathematical model based on the probability of different risks and the effectiveness of different countermeasures, it is also based on psychological reactions of the user to both risks and countermeasures. One example is the use of certified seals in online stores to establish trust and make customers feel safe. The human factor has an effective role in computer security domain, and it is necessary to study and focus on the human side of security issues. The effort and solutions in this step will be added to those found in the technical and formal sides to increase and enhance security levels throughout the system.

The field of HCI provides many methods to study human behavior and interactions with computers. Many methods rely on exploring human cognitive processes that lead to understanding human intelligence and behavior, such as how the human mind works, and how people perceive, remember, think, and learn knowledge. In general, human cognition refers to "The mental action or process of acquiring knowledge and understanding through thought, experience, and the senses" [1].

Metaphors and mental models are common methods for studying human behavior in HCI area. Metaphors are frequently used in user interface (UI), while the mental model interprets human behavior and reasoning, showing how people think and process knowledge. It's an internal representation of external reality that is a conceptual representation formed by the mind. It is used to anticipate events and shape human reasoning and decisions [2]. The primary base of mental model techniques depends on the notion of concept, which is a collection of essential building blocks of human thoughts and beliefs [3]. The mental model portrays human cognitive processing and reasoning via a representation of concepts. The most influential concepts are the main driving force for users' reasoning and decisions.

The human factor in security domain is its correlation with the usability element. Despite the usefulness of security policies, most often users consider the security as a secondary task, especially when they feel it hinders them from completing their primary tasks [4]. Therefore, they desire the security tasks to be quick and easy. When people feel that their primary task will be hindered by security, they resort to finding ways to circumvent security efforts. It is apparent how security is deeply tied up with usability. The mismatch between security goals and usability goals in users' mental models contributes to making inappropriate security-related decisions. Security goals include confidentiality, integrity, and availability, which provide protection for the data and resource. Meanwhile, usability goals concentrate on users' preferences such as effectiveness, efficiency, and satisfaction.

Kainda et al. [5] indicate 6 categories in the current HCISec studies that show where the security issues resulting from human behavior are related to usability. The categories are Authentication, Encryption, PKI (Public Key Infrastructure), Device pairing, Security tools and Secure systems. These categories were determined after extensively reviewing the current HCISec and usability research to identify the main factors of security and usability. Their study concluded that the most security and usability problems appear under one of the 6 aforementioned categories. This could give an extensive understanding of factors affecting human decision-making when it comes to security.

The aim of this paper is to understand users' decision-making in security and usability through the mental model approach. To elicit and depict users' security and usability mental models, crowd sourcing techniques and a cognitive map method are applied. Participants are recruited via MTurk. Sets of questionnaires are designed and distributed on MTurk. The questionnaires are designed based on the categories of [5] for the security and usability measurement. The results and the analysis of the PKI category are also presented. A cognitive map is used to extract and visualize concepts from the participants' answers.

## II. HUMAN FACTOR IN SECURITY

Human factors play a significant role in cyber security. Human behavior impacts both information security and (ultimately) associated risks. For instance, the lack of awareness on cyber threats and failure to comply with security policies have appeared as a user decision that impacts the security of an organization. The weakness of human behavior is becoming an important source for attackers. It shifts attacker's attention from directly attacking the machine to targeting human vulnerability [6]. For instance, instead of using exhaustive efforts in using application programs to decode encrypted data, an attacker could use a social engineering technique to obtain sensitive information (e.g. a phishing method, where the attacker exploits the victim's trust). A successful phishing attack depends primarily on the weakness in a user's awareness and attention. The study of Dhamija et al. in [7] shows how 90% of users failed to identify phishing websites in a controlled lab experiment due to the lack of attention that resulted from visual deception practiced by the phishers. Although in [7] the primary targets for participants was related to security and users knew in advance they had to perform security tasks, a large proportion of them failed to realize risks and perform basic security checks. Despite the severity of human weaknesses in security and protection, the human role has received little attention in security domains. The role of humans is often neglected in favor of technical solutions in the cyber security equation. The technology side is an essential part in cyber security, but people are strongly involved in the information security. They are responsible for the design, implementation, and operation of the technology and tools used in security measures [8]. Preventing technical problems is not the only effective way to assist in maintaining the security of systems. The security of any system depends on human behavior. Recently, human behavioral issues have created more challenging issues to cyber security than more technical issues [9]. Therefore, understanding human factors are vital step in cyber security.

## III. SECURITY AND USABILITY FACTORS

Kainda et al. [5] established the categories of security and usability measurements after reviewing current HCISec and usability research. They then identified the main factors that were the target of measurement for both usability and security. Their study concluded that most security and usability problems appear under one of the following categories:

1) *Authentication*: This refers to affirming the identification of the end user before allowing access to system resources. The focus in this category is to measure: a) *memorability:* the degree to which users can remember the password, or *cognition:* the process of verifying the identity by asking questions about knowledge that only the user should know and b) *efficiency*: the successful authentication within an acceptable amount of time and effort.

2) *Encryption*: the focus is on the users' understanding the mechanisms of sending email securely. Email encryption may overlap with the authentication category. The difference between the two is that, unlike authentication mechanisms that focus on memorability and cognition, email encryption focuses on how users understand the knowledge behind the mechanisms of sending the email securely (e.g. guidelines and rules that assist in avoiding phishing scams emails or awareness tips for downloading attachment files via email).

3) *PKI (Public Key Infrastructure)*: PKI centers around the notion of determining the identity via system interface. The main purpose behind it is to facilitate the transfer of information in a secure manner that users can notice it directly through the system interface. The major focus is on users' knowledge of particular indicators and signs to see if users can distinguish whether the website is secure or not. PKI includes a variety of indicators ranging from the traditional padlock symbol appearing on the bottom of the internet browser, to using different colors in the address bar of the browser. The factors in this category involve vigilance, attention and awareness.

4) *Device pairing*: The focus is on: a) efficiency: the successful attempt of pairing device within an acceptable amount of time and effort, b) *effectiveness* (failure/success of pairing), and c) *security failures*: the process of device pairing is accomplished correctly but with an unfamiliar device. "Security failures" is an independent factor that does not result from inefficient or ineffective pairing. For instance, the user could pair the device successfully within an acceptable amount of time and effort but with the unfamiliar/untrusted device.

5) *Security tools*: Related security tools provide the system safeguards and protection (e.g. antivirus software, password managers, and firewalls). Like encryption, this category focuses on users' knowledge related to using these tools. Lack of user awareness in using the security tools has a negative impact on the effectiveness of these tools.

## IV. SECURITY MENTAL MODEL AND COGNITIVE MAP

### A. Security mental models

A mental model (or mental map) is a "small-scale model" to explain people's thought processes [10]. It represents how people understand ideas and concepts and how they connect these concepts by relations. It is a kind of internal representation of external reality.

The mental model was introduced in the psychology field by Kenneth Craik [11]; he suggested that the human mind builds and constructs "small-scale models" to anticipate events. Later, the mental model is mentioned in other domains like cognitive field where Johnson-Laird proposed the mental model to be a basic structure for cognition. Johnson-Laird stated, "It is now plausible to suppose that mental models play a central and unifying role in representing objects, states of affairs" [12]. The mental model approach in security people's perceptions and cognition of the concepts behind their security decision. Ideally, if people's mental models aligned with security policies, then the security issues and problems would be fewer due to the people's security awareness and perception. However, mental models vary among individuals and depend on the special characteristics and experiences of each person, e.g., a security expert's mental model will be more rich and extensive than a novice user [13].

Several attempts have been made to understand human mental models relating to computer security. Wash [13] identified eight mental models of security threats. His models targeted non-expert computer users. He interviewed people about computer security threats to understand how their thinking leads to security practices like ignoring security advice. Camp [14] proposed five possible high-level mental models for computer security failures in the following areas: physical security, medical infections, criminal behavior, economics failure, and warfare. Her models show how people think about computer security in each of these areas. Coopamootoo et al. [15] used the cognitive map approach to identify privacy concerns and behavior.

*B. Cognitive map*

A cognitive map is a type of mental to enable the analyzing of tasks requiring mental operations [16]. The cognitive map was on the personal construct theory [17] that was originally proposed to understand how humans make sense of and operate in their world. Operational researchers have demonstrated the uses of the cognitive map technique for working on a variety of different tasks. One such task is assisting the construction of complex data in the domain of problem solving where the problem is divided into smaller task sand each task has a particular characteristic to be solved. So, in case of a failure in solving the problem, the cognitive map allowed to pinpoint the possible sources of failure inherent in a task. Furthermore, it provides aid in managing large amounts of qualitative data from documents by simplifying the presentation of data. It visualizes the data as a network consisting of nodes and arrows that represent the type of relationship between the entities of the data and the causal effects. Graph theory can be used to analyze the structure of the cognitive maps and give an explanation on how the cognitive map represents concepts. For instance, graph density indicates how connected or sparse the maps are. If the density of the map is high, people see a large number of causal relationships among concepts. To construct cognitive map there is a variety of existing methods [18] ranging from a) questionnaires, (b) extraction from written texts, (c) drawing a cognitive map from data that shows causal relationships, and (d) through interviews with people who draw it directly.

## V. METHODOLOGY

Our methodology used Amazon's Mechanical Turk and the cognitive mapping technique to elicit and depict users' security and usability mental models. To obtain the cognitive map, the method of extracting the concepts from written text was used. The mental model was constructed by analyzing peoples' answers through survey questionnaires distributed in MTurk. The crowdsourcing practice of recruiting participants via MTurk incorporates crowdsourcing in the security decision-making analysis and to capture people's mental models in a scalable manner. The MTurk platform addresses several problems related to traditional surveys methods. For instance, it speeds up the time process of collecting data where the survey requester has access to a global, on-demand, 24/7 workforce around the world. It also overcomes the problem of answers' quality by offering a clear monetary benefit for participants to complete their task. At the same time, participants know that the requester already has the choice of assessing the quality of their answers and can reject the participant's results if deemed inadequate or neglected. Thus, participants are eager to provide careful answers and pay attention to the tasks.

*A. Design*

Our experiment was a between-subject study that includes a different framing of security and usability related questions requiring free text responses. The set of questions was designed as a Human Intelligence Task (HIT) in AMT. Focus questions centered on solving a particular problem were asked. The expected answer from participants was unstructured text and the text was analyzed to extract the main concepts used to construct the mental model. The questionnaire list had 5 sections each representing one security and usability category from [5]. Participants were equally divided and assigned randomly to a section. About 232 participants were recruited. Due to space constraints of this paper, the results and the analysis of one group, the PKI group (46 participants), are presented. Table 1 shows the demographic information for participants including gender, age range, and education. The demographic information may be used for further investigation. For instance, does the education level usually reflect people's behavior and attitude, and would this impact their decision-making on security? Also, can age range associated with peoples' experience influence the security decision-making?

To guarantee the quality of the data, participants in this study were required to have a lifetime MTurk approval rate (this refers to the rate of successfully completing previous tasks) greater than 50%. HITs were completed over 3 days. 46 responses were received, but due to incomplete answers 2 were discarded, yielding 44 valid responses. The average time that participants spent per HIT was 11.20 minute (M=75.07, SD=54.13), and they were paid at the rate of $1 per HIT.

*B. Text process and visulazation*

We used CMapTools* to develop and visualize the concept maps of the mental model. The hierarchy structure of concepts offered a structured representation of a participant's conceptual understanding security and usability. The map structure is a directional graph consisting of nodes and edges: each node labels a distinct concept, while edges depict the relationships between concepts and portray participants' thought processes toward a particular subject. The direction of the edge represents the cause/effect or means/ends relation. After collecting the free text, the following process for each response was performed:

**Table 1**- Number of participants based on age, gender and educational level

| Male | %58.69 | | Female | %41.30 |
|---|---|---|---|---|
| Age | Under 18 years | 0 | 18 to 24 years | 6 |
| | 25 to 34 years | 19 | 35 to 44 years | 15 |
| | 45 to 54 years | 3 | 55 and older | 3 |
| Education | Below high school | 0 | High school | 3 |
| | Some college level | 7 | Associate degree | 5 |
| | Bachelor's degree | 21 | Some postgraduate | 1 |
| | Master's degree | 9 | PhD and above | 0 |

---
\* https://cmap.ihmc.us/

1- Create two sets:
   a) *Concept_set*: set of concepts in the mental model.
   b) *Delete_set*: the deleted content from the text.
2- Preprocess the text:
   a) Clean up the text: Check spelling, correct typos and move the non-content words from text to the Delete set. E.g. conjunctions, noise words, numbers... etc.
   b) Normalize words in the text by detecting word derivations and replacing them with the normal forms.
   c) Identify synonyms: words that have the exact or similar meaning to another word are combined.
3- Divide the text into distinct sentences.
4- For each sentence in the text starting from the first one:
   a) Identify and classify concepts and relations such as concept A performing an action on object concept B.
   b) If concept A does not exist in the *Concept_set*, add concept A to *Concept_set* and a vertex labeled with concept A is created. Similar situation for concept B. Arrow from A to B is labeled with identified relation.
   c) Repeat steps a and b for the entire sentence in the text.

The previous steps produced a mental model for each response by each individual. To develop a composite conceptual map involving all participants the individual concept maps were aggregated into one map for each question. The aggregation process unified the sets of nodes in the *Concept_set* for all individuals' maps with its relations. Each edge in the graph is labelled by the number of compatible responses form participants. For instance, Figure 1.b shows 5 participants relating the concept "look" with the concept "professional". Due to space limitations we only display the concepts relating to security and usability.

*C. Result and analysis*

The *PKI* category involved 2 questions that tested participant security vigilance, attention and awareness. The first question aimed to see if participants could distinguish if the web page URL in the address bar was secure or not. The security of the URL could be identified quickly by the padlock symbol in the address bar where the style and the color of the URL changed depending on the security status of the website. In Figure 2 (see Appendix), screenshots from four browsers are presented as online car stores, and the participants were asked to choose which store they preferred to use in order to buy a car. The intent of the first question was to not mention "security," "trust," or any other word that could shift participants' minds to security. Instead the process of buying a car was specified because to see if a buyer would still think of safety and buying from a reputable source. Each URL in the screenshot had different style. The following shows the format of each choice with the percentage of participants' choices:

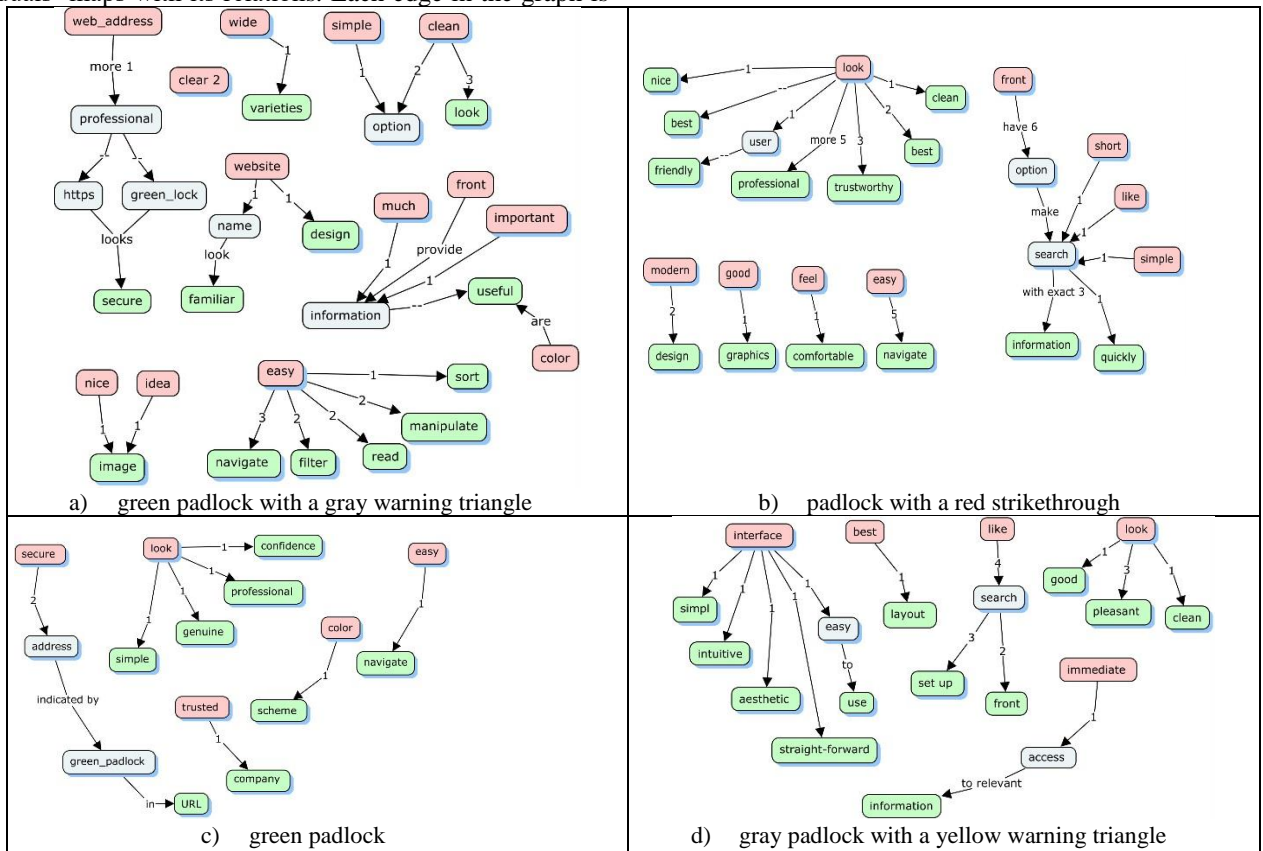

a) green padlock with a gray warning triangle
b) padlock with a red strikethrough
c) green padlock
d) gray padlock with a yellow warning triangle

**Figure 1 -** Cognitive maps of question one

a) green padlock with a gray warning triangle: the site is secure but browser had blocked insecure content. (36.36%)
b) padlock with a red strikethrough: the connection is not secured. (36.36%)
c) green padlock: the connection is encrypted. (9.09%)
d) gray padlock with a yellow warning triangle: the connection is only partially encrypted and doesn't prevent some security issues. (18.18%) only a few made the safest choice (green padlock), and of those only 2 individuals chose this option for security reasons. The mental map of the green padlock option Figure 1.c shows 2 participants indicated that the website "address" is "secure" as indicated by the "green_padlock" in the "URL". Other participants' justifications belonged to usability concepts such as "easy" to "navigate," "look," "genuine," "professional," and "confidence". Security vigilance also appeared in the mental map of option a (green padlock with a gray warning triangle) Figure 1.a, where one person also pointed out that the web address is more "professional" and the "https" "green lock" looks "secure". The rest of the concepts in the map belong to usability concepts. In the concept map of choice b (padlock with a red strikethrough), 3 participants indicated the concept of "trustworthy" and linked it directly with the concept "look", which indicates that the website general appearance and interface is the trigger point for participants' trustworthiness. All other concepts in the choice map belong to the usability element.

The second question aimed to was to test participants' vigilance and attention in mobile applications. Similar to question 1, a set of Android and Apple app screenshots were presented and participants were asked to choose one app to download as a security app. In this question, participants were divided into two sections based on the type of their cellphone. Contrary to the first question, the word security was explicitly mentioned in the question to see if this could influence participants' decisions. The criteria that we used to edit the screenshots of the Android group are: a) *people review*: the reviews are considered to be personal recommendations, usually establishing the factor of trust in people and affecting their behavior; b) *app ads*: advertisement shows in apps; c) *top developer sign*: a sign provided by *playstore* from Google to indicate the app quality; and d) *a fake green thumb symbol* (originally, the green thumb symbol did not exist in the Apple app stores and thus does not have meaning, but it is presented in the screenshot to see if it has an effect on participants' decisions). Surprisingly the choice of the fake symbol Figure 3.a (see Appendix) received the highest percentage of choices (72.73%). The concept map of this choice Figure 4.a (see Appendix) shows that the factor with the largest effect on participants' decisions was people's "great" rating where 14 participants relate the concept "great" with "rating". Not having ads was the second factor. The concept "safe" appeared as a receiver node with the degree of 3. The trigger concepts responsible for the appearing of concept "safe" are "feel", "lots", "downloads", "contain" and "ads" were found by tracing the direct and indirect connected nodes. People's feelings, the number of downloads indicated by people's reviews, and adds are the factors behind indicating the concept "safe". Also, the concept "feel" related directly with the concept "trust". None of the participants in the Android section noticed the formal symbol of top developer (only one participant chose this option and the mental map shows the concept "nice" to describe the reason behind choosing it).

In the Apple group, we used the following criteria to edit the apps' screenshots: a) people review; b) green thumb symbol (like in the Android group); and c) age tag highlighting the recommended age. Similar to the Android group, the green thumb option received the highest rate of selection (41.67%). Figure 4.b show the concept map of this option. The map shows 5 clusters of participants' justifications. In the thumb symbol cluster, participants linked the concept "thumbs up" with the following concepts: "reassure," "positive," "rating," and "liked." The thumbs up symbol gave a positive impression for the app and participants thought it was related to other people's positive rating. Although the same app rating is presented under the app name in the most options, people were more influenced by the green thumb symbol.

## VI. CONCLUSION AND FUTURE WORK

The mental model approach provides a valuable framework to explore the links between security decision-making and usability. It illustrates the cognitive processing and reasoning behind the end user decision by depicting the concept set that leads the end user to make a specific decision. Exploring the security knowledge and cognitive processes of people will assist researchers in creating security policies and designing and implementing security tools. Thus, contributing towards more usable security. This paper presents a composite mental model for peoples' decision-making toward security and usability. To construct the model, the practice of crowd sourcing technique and cognitive map method were used. The cognitive map was based on the text analysis method. A set of questionnaires in the categories of the security and usability [5] was designed and distributed via MTurk. The text responses were then collected and analyzed. Afterwards, the results and analysis of the PKI category were presented. The cognitive map was used to extract and visualize the concepts from the participants' answers. The PKI result shows that security vigilance, attention and awareness were low due to the usability factor (e.g. in question 1 only 2 participants (9.09%) were aware of the safest URL option in the browser, while the attention of the other participants shifted to usability factors relating to the websites' general appearances and interfaces. The analysis also shows that participants' decisions were influenced by the green color (green thumb symbol and the green app logo). The screenshots including the green color in all questions were the ones most selected by participants. Moreover, the concept "feel" frequently appeared in some cognitive maps which is agrees with the findings in [19] where it shows how security decisions are affected by people's feelings.

For future work, we would like to expand the security mental model and explore the cognitive map of users in other security and usability categories. Also, since the mental model methodology resulted in producing a graph, graph theoretical indices were used to analyze the structure of the cognitive maps. For instance, concept centrality gives a valuable insight for how important the concept is to the other concepts, as well as which were the most important concepts that affected concepts.

## Appendix A

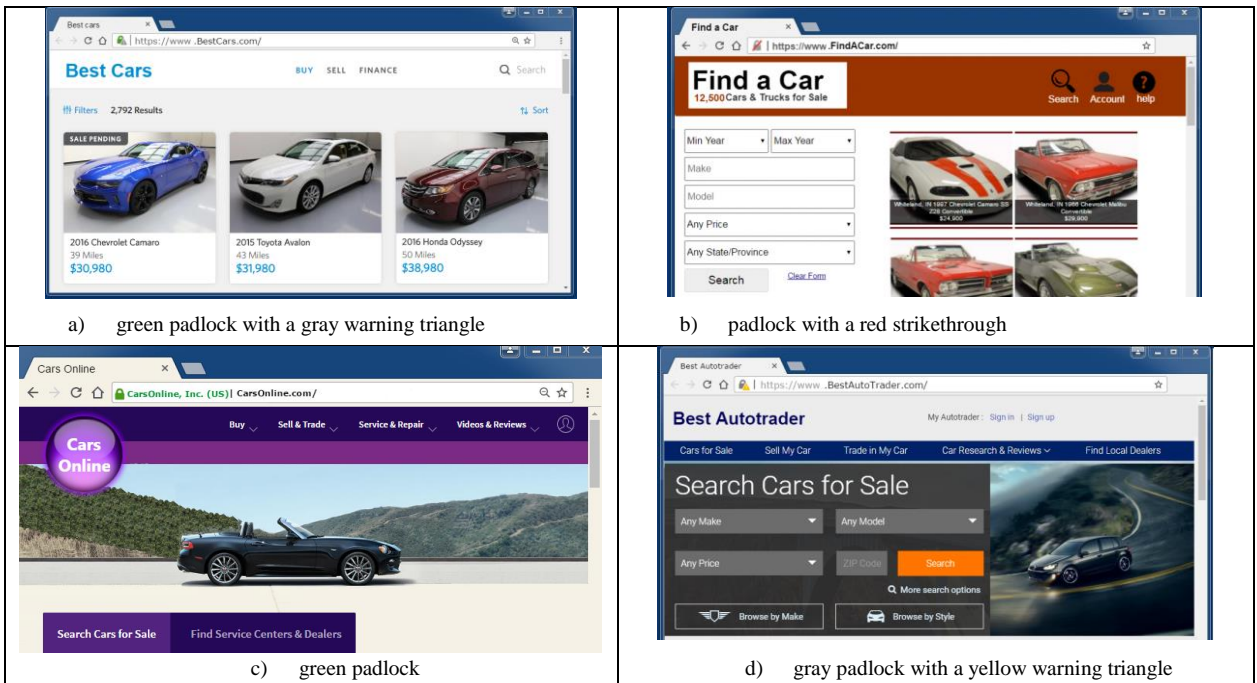

a) green padlock with a gray warning triangle
b) padlock with a red strikethrough
c) green padlock
d) gray padlock with a yellow warning triangle

**Figure 2 -** Browsers' screenshots for cars online stores

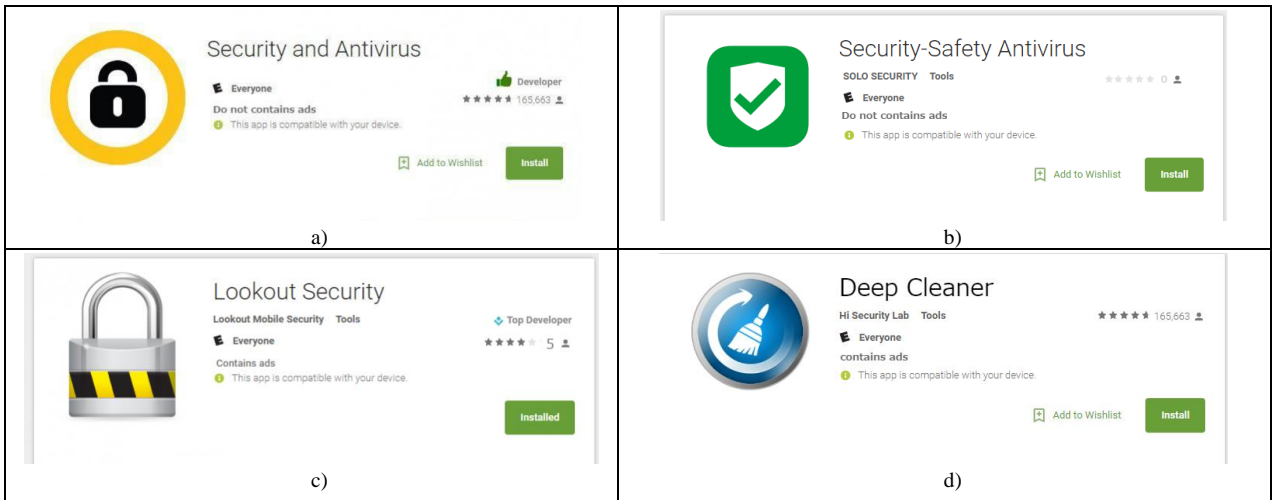

Figure 3 - Android app screenshots

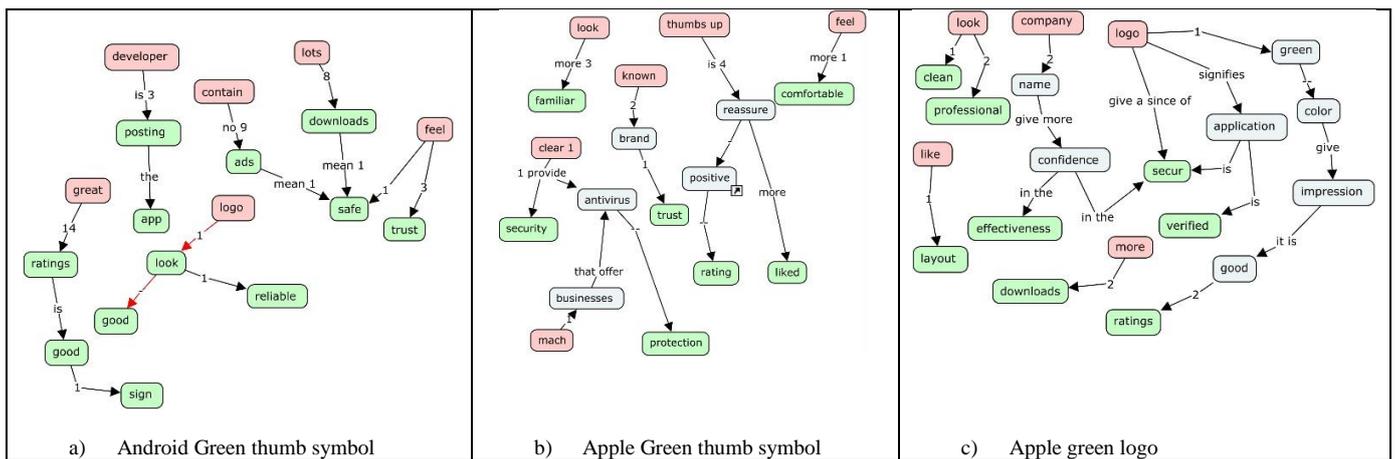

a) Android Green thumb symbol     b) Apple Green thumb symbol     c) Apple green logo

Figure 4 – Cognitive map of Apple